\documentclass[12pt,letter]{article}
\pdfoutput=1
\usepackage{graphicx, epsfig, color,cite}
\usepackage{amsmath}
\usepackage{amssymb}
\usepackage{caption,subcaption,graphicx}
\usepackage[hidelinks]{hyperref}
\def\eslt{\not\!\!\!{E_T}}
\def\to{\rightarrow}

\def\bi{\begin{itemize}}
\def\ei{\end{itemize}}

\def\tst{\tilde t}

\def\tg{\tilde g}

\def\tell{\tilde\ell}
\def\tq{\tilde q}

\def\tw{\widetilde\chi^{\pm}}
\def\tz{\widetilde\chi^0}
\def\alt{\lesssim}
\def\agt{\gtrsim}
\def\be{\begin{equation}}  
\def\ee{\end{equation}}  
\def\bea{\begin{eqnarray}}  
\def\eea{\end{eqnarray}}

\begin{document}
\begin{titlepage}
\begin{flushright}
OU-HEP-190131
\end{flushright}

\vspace{0.5cm}
\begin{center}
{\Large \bf LHC SUSY and WIMP dark matter searches\\
confront the string theory landscape
}\\ 
\vspace{1.2cm} \renewcommand{\thefootnote}{\fnsymbol{footnote}}
{\large Howard Baer$^1$\footnote[1]{Email: baer@ou.edu },
Vernon Barger$^2$\footnote[2]{Email: barger@pheno.wisc.edu},
Shadman Salam$^1$\footnote[3]{Email: shadman.salam@ou.edu},\\
Hasan Serce$^2$\footnote[4]{Email: serce@ou.edu}
and 
Kuver Sinha$^1$\footnote[5]{Email: kuver.sinha@ou.edu}
}\\ 
\vspace{1.2cm} \renewcommand{\thefootnote}{\arabic{footnote}}
{\it 
$^1$Dept. of Physics and Astronomy,
University of Oklahoma, Norman, OK 73019, USA \\[3pt]
}
{\it 
$^2$Dept. of Physics,
University of Wisconsin, Madison, WI 53706 USA \\[3pt]
}

\end{center}

\vspace{0.5cm}
\begin{abstract}
\noindent
The string theory landscape of vacua solutions provides physicists with some understanding 
as to the magnitude of the cosmological constant. Similar reasoning can be applied to the
magnitude of the soft SUSY breaking terms in supersymmetric models of particle physics: 
there appears to be a statistical draw towards large soft terms which is tempered by the 
anthropic requirement of the weak scale lying not too far from $\sim 100$ GeV. For a mild statistical
draw of $m_{soft}^n$ with $n=1$ (as expected from SUSY breaking due to a single $F$ term) then
the light Higgs mass is preferred at $\sim 125$ GeV while sparticles are all pulled beyond LHC bounds.
We confront a variety of LHC and WIMP dark matter search limits with the statistical expectations
from a fertile patch of string theory landscape. The end result is that LHC and WIMP dark matter detectors
see exactly that which is expected from the landscape: a Standard Model-like Higgs boson of mass
125 GeV but as yet no sign of sparticles or WIMP dark matter. SUSY from the $n=1$ landscape
is most likely to emerge at LHC in the soft opposite-sign dilepton plus jet plus MET channel.
Multi-ton noble liquid WIMP detectors should be able to completely explore the
$n=1$ landscape parameter space.

\end{abstract}
\end{titlepage}

\section{Introduction}
\label{sec:intro}

\begin{figure}[tbp]
\begin{center}
\includegraphics[height=0.3\textheight]{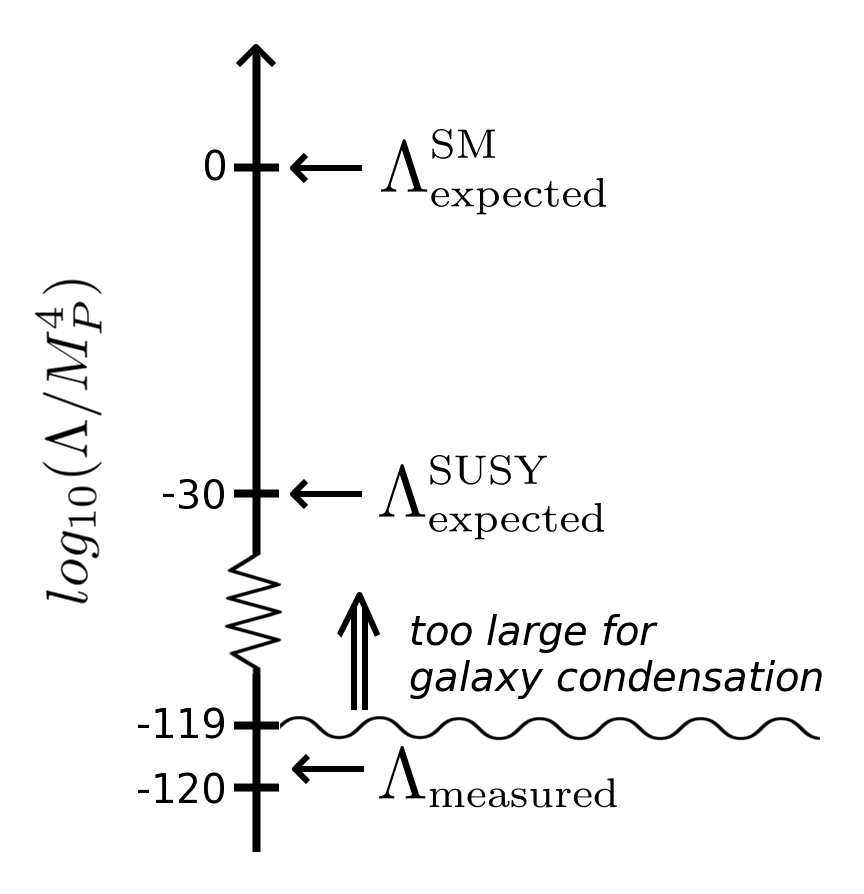}
\caption{
Log portrayal of expected parameter space of the cosmological constant $\Lambda$
from the string theory landscape.
\label{fig:CC}}
\end{center}
\end{figure}

It is sometimes lamented that the emergence of the landscape of string 
vacua \cite{Bousso:2000xa,Susskind:2003kw,Bousso:2004fc}
has rendered string theory non-predictive since how are we to pick out the
(meta-stable) vacuum corresponding to our universe from (perhaps) of order
$10^{500}$ possibilities? 
Such sentiment ignores one of the great predictions of the 
latter 20th century \cite{Weinberg:1987dv}: 
namely that given a multiverse which includes a vast 
assortment of pocket-universes with varying cosmological constants, then
it may not be surprising to find $\Lambda\sim 10^{-120}m_P^4$ since if 
it was much bigger, then galaxy condensation would not occur and we 
would not even be here to discuss the issue.
The situation is portrayed in Fig. \ref{fig:CC} which depicts the 
fact that the cosmological constant ought to be at its most natural value
{\it subject to the constraint that we can exist to observe it}. 
Such anthropic reasoning relies on the existence of a vast landscape of 
possibilities that is provided for by the discretuum of flux vacua from 
string theory \cite{Bousso:2000xa,Susskind:2003kw,Bousso:2004fc,Douglas:2003um}. 

Can such reasoning be applied to the origin of other mass scales 
that appear in fundamental physics?\footnote{Weinberg states \cite{Weinberg:2005fh}: 
``The most optimistic hypothesis
is that the only constants that scan are the few whose dimensionality is a positive power of mass:
the vacuum energy, and whatever scalar mass or masses set the scale of electroweak
symmetry breaking.''} 
An obvious target would be the magnitude of the weak scale 
($m_{weak}\simeq m_{W,Z,h}\simeq 100$ GeV) and why it is
suppressed by 16 orders of magnitude compared to the (reduced) 
Planck scale $m_P\simeq 2.4\times 10^{18}$ GeV. Anthropically,
one expects a universe with electroweak symmetry properly broken
such that weak bosons gain mass $m_{W,Z}\sim 100$ GeV while
solutions with charge or color breaking minima would be excluded.
In addition, nuclear physics calculations by 
Agrawal {\it et al.} \cite{Agrawal:1997gf}
require that the magnitude of the weak scale shouldn't exceed its measured
value by a factor of 2-3 in order to gain a livable universe. 

Given that quantum corrections to the Higgs mass diverge quadratically
with the theory cutoff $\Lambda_{SM}$, 
it seems the Standard Model (SM) with $\Lambda_{SM}\gg m_{weak}$ 
would be a rare occurrance  within the landscape since one would be 
required to select only those vacuum solutions with highly fine-tuned
scan parameters. In a landscape containing both SM and supersymmetric SM
solutions (SSM), one would expect vastly more SSM solutions with 
$m_{weak}\simeq 100$ GeV since then quantum corrections to $m_h$ diverge only
logarithmically. 

In fact, this discussion brings up the issue of how naturalness is connected
to vacua selection in the multiverse. 
Here we adopt the notion of {\it practical naturalness}: 
\begin{quotation} 
An observable ${\cal O}\equiv o_1+\cdots +o_n$ 
is natural if all {\it independent} contributions $o_i$ 
to ${\cal O}$ are comparable to or less than ${\cal O}$. 
\end{quotation}
This is because if any $o_i\gg {\cal O}$, then some other contribution $o_j$ ($j\ne i$)
would have to be fine-tuned to a large opposite-sign value to keep ${\cal O}$ at
its measured value. 
The notion of practical naturalness may be compared with what Douglas calls 
stringy naturalness \cite{Douglas:2004zg}:
\begin{quotation}
An effective field theory (or specific coupling or observable) $T_1$ is more natural in string
theory than $T_2$ if the number of {\it phenomenologically acceptable} vacua leading to $T_1$ is larger than the
number leading to $T_2$.
\end{quotation}
In a landscape of vacua where independent contributions $o_i$ to observable $\cal O$ are 
uniformly distributed, then it follows that many more vacua are likely to exist where
the $o_i$ are comparable to ${\cal O}$ than where some $o_i\gg{\cal O}$ so that
some other value $o_{j\ne i}\simeq -o_i$. Thus, we expect these two definitions to be
equivalent descriptions of naturalness.
The landscape, if it is to be predictive, is predictive in the 
statistical sense: the more prevalent solutions are statistically more likely.
This gives the connection between landscape statistics and naturalness: 
vacua with natural observables are expected to be far more common than vacua
with unnatural observables.\footnote{Since the landscape allows for an apparently unnatural 
value of $\Lambda_{cc}\ll m_P^4$, it is sometimes interpreted 
that vacua with {\it other} highly unnatural observables should also be 
entertained.
In the case of $\Lambda_{cc}$, we should find ourselves in a universe where
the cosmological constant is {\it as natural as possible} such that we gain a 
livable universe. Then we would expect $\Lambda_{cc}\sim 10^{-120}m_P^4$ 
rather than say $10^{-200}m_P^4$.}

Thus, in this paper we will focus on vacua solutions which include the 
Minimal Supersymmetric Standard Model (MSSM) as their weak scale 
effective theory. 
This restricts our attention to a {\it fertile patch} of landscape 
solutions which include the SM as the weak scale effective theory
(in accord with experiment) but where the weak scale is stable against 
quantum corrections (as in the MSSM). 
We will further assume that the MSSM arises from 
a 4-d supergravity theory (SUGRA) where SUSY breaking occurs via spontaneous
SUGRA breaking in a hidden sector of the theory with a, perhaps, complicated
SUSY breaking sector including possibly numerous SUSY breaking
$F$- and $D$-term fields which gain vevs. 
The question that can be addressed is then: 
what does a statistical sampling of this fertile patch of 
the landscape say about the scale of SUSY breaking, and hence the 
likelihood of observable signals at the CERN Large Hadron Collider (LHC) or
at WIMP dark matter direct and indirect detection experiments?
Indeed, this question has already been investigated early on by
Douglas {\it et al.} \cite{denefdouglas,doug1,doug2}, Susskind \cite{suss} 
and by Dine {\it et al.} \cite{dine}. 
(For some reviews, see {\it e.g.} Ref's \cite{DK,Kumar:2006tn}.)

\section{Statistics of the SUSY breaking scale}

In this Section, we assume a vast ensemble of string vacua states which give 
rise to a $D=4$, $N=1$ supergravity effective field theory at high energies. 
Furthermore, the theory consists of a visible sector containing the MSSM along 
with a perhaps large assortment of
fields that comprise the hidden sector. 
The scalar potential is given by the usual supergravity form \cite{Nilles:1983ge}
\bea
V&=&e^{K/m_P^2} \left( g^{i \overline{j}} D_{i}W D_{\overline{j}}W^{*} \, - \, 
\frac{3}{m_P^2} |W|^2 \right) \, + \, \frac{1}{2}\sum_\alpha D^2_\alpha \,\, \\
&=&e^{K/m_P^2} \left( \sum_i|F_i|^2-3\frac{|W|^2}{m_P^2}\right)+\frac{1}{2}\sum_\alpha D_\alpha^2
\eea
where $W$ is the holomorphic superpotential, $K$ is the real K\"ahler potential
and $F_i=D_iW=DW/D\phi^i\equiv\partial W/\partial\phi^i+(1/m_P^2)(\partial K /\partial\phi^i)W$
are the $F$-terms and $D_\alpha\sim \sum\phi^\dagger gt_{\alpha}\phi$ are the $D$-terms
and the $\phi^i$ are chiral superfields.
Supergravity is assumed to be broken spontaneously via the super-Higgs mechanism either via
$F$-type breaking or $D$-type breaking or in general a combination of both leading 
to a gravitino mass $m_{3/2}=e^{K/2m_P^2}|W|/m_P^2$. 
The (metastable) minima of the scalar potential can be found by requiring
$\partial V/\partial\phi^i=0$ with $\partial^2V/\partial\phi^i\partial\phi^j>0$ to ensure a local
minimum.
The cosmological constant is given by
\be
\Lambda_{cc} \,\, = \,\, m_{hidden}^4\, - \, 3 e^{K/m_P^2}|W|^2/m_P^2 \,\, 
\ee
where $m_{hidden}^4=\sum_i |F_i|^2 \, +  \, \frac{1}{2}\sum_\alpha D^2_\alpha $ is a mass scale associated
with the hidden sector (and usually in SUGRA-mediated models it is assumed $m_{hidden}\sim 10^{12}$ 
GeV such that the gravitino gets a mass $m_{3/2}\sim m_{hidden}^2/m_P$).

According to Douglas {\it et al.} \cite{doug1} from investigations of flux compactifications in 
IIB string theory, the distribution of vacua ought to have the form
\be
dN_{vac}[m_{hidden}^2,m_{weak},\Lambda ]=f_{SUSY}(m_{hidden}^2)\cdot f_{EWFT}\cdot f_{cc}\cdot dm_{hidden}^2
\ee
where we define the weak scale $m_{weak}\simeq m_{W,Z,h}\simeq 100$ GeV and where
$m_{hidden}$ sets the scale for SUSY breaking with $m_{hidden}^2=\sum_i|F_i|^2+\frac{1}{2}\sum_{\alpha}D_{\alpha}^2$
for a (in general) more complicated SUSY breaking sector containing multiple sources of SUSY breaking,
as may be  expected to occur in string theory. 

The function $f_{SUSY}$ contains the expected statistical
distribution of SUSY breaking scales. This is related to the mass scale of MSSM soft terms as
$m_{soft}\simeq m_{hidden}^2/m_P$. If the sources of SUSY breaking have uniformly distributed 
vacuum expectation values (vevs), then it is surmised that
\be
f_{SUSY}(m_{hidden}^2)\sim (m_{hidden}^2)^{2n_F+n_D-1}
\ee
where $n_F$ is the number of $F$-breaking fields and $n_D$ is the number of $D$-term breaking
fields in the hidden sector\cite{denefdouglas,doug1,suss,Douglas:2004zg}. 
We will denote the collective exponent in $f_{SUSY}$ as $n\equiv 2n_F+n_D-1$. 
Since the $F$ terms are complex-valued but the modulus $|F|$ sets the scale of SUSY breaking, 
then they contribute as $(m_{hidden}^2)^{2n_F}$ whereas the real valued $D$ terms contribute
as $(m_{hidden}^2)^{n_D}$. In terms of MSSM soft SUSY breaking parameters, one would
expect a statistically {\it uniform} distribution of soft terms $m_{soft}^0$ only for a single
$D$-term breaking field so that $n_D=1$. A single $F$-term breaking field leads to $f_{SUSY}\sim m_{soft}^1$
so that there is a linearly increasing preference for large soft terms. For more complex configurations
with larger number of $n_F$ and $n_D$, then there is an even greater statistical preference for
large soft terms which could lead to a preference for models with high scale SUSY breaking.

Regarding the role of the cosmological constant in determining the SUSY breaking scale, 
a key observation of Denef and Douglas \cite{denefdouglas,doug1} and 
Susskind \cite{suss} was that $W$ at the minima is distributed uniformly as a complex variable, and the distribution of 
$e^{K/m_P^2}|W|^2/m_P^2$ is not correlated with the distributions of $F_i$ and $D_\alpha$. 
Setting the cosmological constant to nearly zero, then, has no effect on the distribution of 
supersymmetry breaking scales. 
Physically, this can be understood by the fact that the superpotential receives contributions 
from many sectors of the theory, supersymmetric as well as non-supersymmetric. 
The cosmological fine-tuning penalty is $f_{cc}\sim \Lambda/m^4$ where the above discussion
leads to $m^4\sim m_{string}^4$ rather than $m^4\sim m_{hidden}^4$, 
rendering this term inconsequential for determining the number of vacua 
with a given SUSY breaking scale.

The final term $f_{EWFT}$ merits some discussion. 
Following Ref. \cite{ArkaniHamed:2004fb}, 
an initial guess \cite{doug1,suss,dine} for $f_{EWFT}$ was that 
$f_{EWFT}\sim m_{weak}^2/m_{soft}^2$ which
may be interpretted as conventional naturalness in that the larger the Little Hierarchy
between $m_{weak}$ and $m_{soft}$, then the greater is the fine-tuning penalty.
As pointed out in Ref. \cite{Baer:2017uvn}, there are several problems with this ansatz.
\begin{enumerate}
\item As soft terms such as the trilinear $A_t$ terms increase, one is ultimately forced
into charge-or-color-breaking vacua of the MSSM \cite{Casas:1995pd,Baer:1996jn}.
These sorts of vacua must be entirely vetoed on anthropic grounds.
\item As high-scale soft terms such as $m_{H_u}^2$ increase too much, 
then they are no longer driven to negative values and electroweak symmetry isn't even broken.
These non-EWSB solutions also should be vetoed on anthropic grounds.
\item As the high scale soft term $m_{H_u}^2$ increases, its weak scale
value actually becomes smaller and smaller until EWSB is barely 
broken \cite{Giudice:2006sn,Baer:2016lpj}.
This means the {\it weak scale } value of $m_{H_u}^2$ becomes more natural-- 
a phenomena known as {\it radiatively driven naturalness} (RNS) \cite{ltr,rns}.
\item As the soft term $A_t$ increases, then cancellations can occur in the 
$\Sigma_u^u(\tst_{1,2})$ contributions to the weak scale scalar potential,
rendering their contributions closer to, not further from, the weak scale 
whilst at the same time lifting up the Higgs mass $m_h$ to the 125 GeV range.
\item Even in the event of appropriate EWSB, the factor 
$f_{EWFT}\sim m_{weak}^2/m_{soft}^2$ penalizes but does not forbid vacua with a weak
scale far larger than its measured value. 
In contrast, Agrawal {\it et al.} \cite{Agrawal:1997gf}
have shown that a weak scale larger than $\sim 3$ times its measured value
would lead to much weaker weak interactions and a disruption in nuclear synthesis
reactions, and likely an unlivable universe as we know it. In addition, Susskind
posits that an increased weak scale would lead to larger SM particle masses and
consequent disruptions in both atomic and nuclear physics. 
From these calculations, it seems reasonable to {\it veto} SM-like vacua which lead
to a weak scale more than (conservatively) four times its measured value.
\end{enumerate} 

To account for these issues, in Ref. \cite{Baer:2017uvn} the alternative of
\be
f_{EWFT}\sim \Theta (30-\Delta_{\rm EW})
\label{eq:fEWFT}
\ee 
was suggested where $\Delta_{\rm EW}$ is the electroweak fine-tuning measure which just 
requires that {\it weak scale} contributions to $m_Z^2$ should be 
comparable to or less than $m_Z^2$. 
From the minimization conditions for the MSSM Higgs potential \cite{WSS} one finds
\be 
\frac{m_Z^2}{2} = \frac{m_{H_d}^2 + \Sigma_d^d -
(m_{H_u}^2+\Sigma_u^u)\tan^2\beta}{\tan^2\beta -1} -\mu^2 \simeq 
-m_{H_u}^2-\Sigma_u^u-\mu^2 .
\label{eq:mzs}
\ee 
The naturalness measure $\Delta_{\rm EW}$ 
compares the largest contribution on the right-hand-side of Eq. \ref{eq:mzs} 
to the value of $m_Z^2/2$.
The radiative corrections $\Sigma_u^u$ and $\Sigma_d^d$ include contributions 
from various particles and sparticles with sizeable Yukawa and/or gauge
couplings to the Higgs sector.
Usually the most important of these are
\be
\Sigma_u^u (\tst_{1,2})= \frac{3}{16\pi^2}F(m_{\tst_{1,2}}^2)
\left[ f_t^2-g_Z^2\mp \frac{f_t^2 A_t^2-8g_Z^2
(\frac{1}{4}-\frac{2}{3}x_W)\Delta_t}{m_{\tst_2}^2-m_{\tst_1}^2}\right]
\label{eq:Sigmat1t2}
\ee
where $f_t$ is the top-quark Yukawa coupling, 
$\Delta_t=(m_{\tst_L}^2-m_{\tst_R}^2)/2+M_Z^2\cos 2\beta (\frac{1}{4}-\frac{2}{3}x_W)$, $x_W\equiv\sin^2\theta_W$, 
$F(m^2)= m^2\left(\log\frac{m^2}{Q^2}-1\right)$ and
the optimized scale choice for evaluation of these corrections is
$Q^2=m_{\tst_1}m_{\tst_2}$.
In the denominator of Eq.~\ref{eq:Sigmat1t2},
the tree level expressions of $m_{\tst_{1,2}}^2$ should be used.
Expressions for the remaining $\Sigma_u^u$ and $\Sigma_d^d$ terms
are given in the Appendix of Ref. \cite{rns}.

In the remainder of this paper, we will assume a solution to the SUSY $\mu$ problem
such as the gravity-safe, electroweak natural axionic hybrid CCK model
based on a $\mathbb{Z}_{24}^R$ symmetry in Ref. \cite{Baer:2018avn}. 
As such, we invert the usual usage of $\Delta_{\rm EW}$ with
fluid soft terms and $\mu$ term: instead, the weak scale (as typified by the value of $m_Z$) is no longer
fixed at its measured value but is instead {\it determined} by Eq. \ref{eq:mzs}. 
In this case, values of $\Delta_{\rm EW}\agt 30$ correspond to a value of 
$m_{weak}\agt$ four times its measured value (in our sub-universe).
The $\Theta$ function in Eq. \ref{eq:fEWFT} guarantees that we veto vacua with CCB minima or
no EWSB. It also vetoes properly broken Higgs potentials but where the weak scale
is generated at more than four times its measured value.

\subsection{Brief review of some previous work and goals of the present work}
\label{ssec:review}

In Ref. \cite{Baer:2016lpj}, an approach similar to Weinberg's anthropic solution 
to the cosmological constant was applied to determination of the
SUSY breaking scale. 
It was assumed that there was a mild draw of the landscape
towards large soft terms which was tempered by the anthropic requirement of
a value for the weak scale which was not too far from its measured value by 
a factor $\sim$ four. The draw of $m_{H_u}^2$ towards large values, tempered
by an appropriate breakdown of EW symmetry, led to {\it barely-broken} 
EW symmetry. This is the same as the naturalness condition that $m_{H_u}^2$
is driven to small negative values at the weak scale, and so gave a {\it mechanism}
for why $m_{H_u}^2$ should be driven to natural values. 
It was also emphasized that the statistical draw to large soft terms 
must avoid CCB and no EWSB vacua while at the same time drawing towards
a weak scale not too far from its measured value. The combined statistical/anthropic draw would pull soft terms towards a region where 
$A_t$ is large (but not too large) and where $m_h\sim 125$ GeV.

In Ref. \cite{Baer:2017uvn}, previous work by DD was adopted wherein
soft SUSY breaking terms were actually selected 
according to $f_{SUSY}\sim m_{soft}^n$ with $n=2n_F+n_D-1$ while
$f_{EWFT}$ was adopted as in Eq. \ref{eq:fEWFT}. This allowed for
{\it landscape probability distributions} to be calculated for a host
of superparticle and Higgs masses. Calculations were performed
using the three-extra parameter non-universal Higgs model (NUHM3) 
with parameter space given by
\be
m_0(1,2),\ m_0(3),\ m_{1/2},\ A_0,\ \tan\beta ,\ \mu,\ m_A\ \ \ (\rm NUHM3)
\ee
where separate first/second and third generation soft scalar masses 
and a negative $A_0$ term were used.
For $n=1$ or 2, then the differential probability distribution for the Higgs
mass $dP/dm_h$ acquired a firm peak around $m_h\sim 125$ GeV.

\begin{figure}[tbp]
\begin{center}
\includegraphics[height=0.3\textheight]{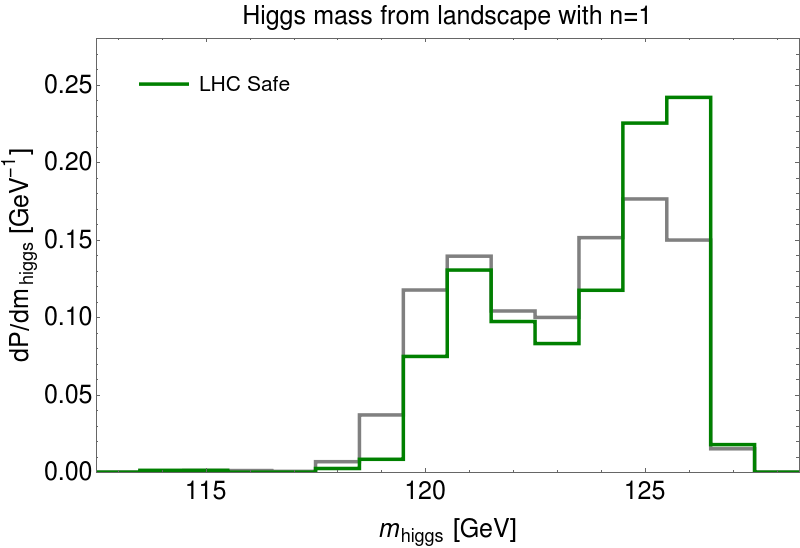}
\caption{
Statistical expectation for the mass of the Higgs boson from
the string theory landscape which scans over single $F$-term SUSY breaking.
The green histogram includes only LHC Run 2 safe points.
\label{fig:mhiggs}}
\end{center}
\end{figure}

Results from a scan over soft terms with $\mu : 100-360$ GeV with $A_0<0$ and 
$n=1$ (corresponding to a single SUSY breaking field $F$) 
are shown in Fig. \ref{fig:mhiggs} for the case where the generated
weak scale is less than a factor four from its measured value.
The green histogram shows results when LHC search constraints (except Higgs mass) are also imposed
(see below).

Furthermore, the probability distributions for other sparticle masses gave
\begin{itemize}
\item $m_{\tg}\sim 4\pm 2$ TeV,
\item $m_{\tst_1}\sim 1.5\pm 0.5$ TeV,
\item $m_A\sim 3\pm 2$ TeV,
\item $\tan\beta \sim 13\pm 7$,
\item $m_{\tw_1,\tz_{1,2}}\sim 200\pm 100$ GeV,
\item $m_{\tz_2}-m_{\tz_1}\sim 7\pm 3$ GeV and
\item $m_{\tq,\tell}\sim 20\pm 10$ TeV.
\end{itemize}
From these results, one may conclude that the present LHC Run 2 results
of finding a SM-like Higgs boson with $m_h=125$ GeV and no sign of sparticles
is seeing exactly that which the landscape with $n=1$ or 2 predicts
will be seen. Furthermore, the higgsino-like WIMPs would form only a portion
of dark matter (along with a SUSY DFSZ-like axion). Since the higgsino-like
WIMPs typically constitute only $\sim 10$\% of dark matter, their 
detection rates lie below present WIMP search limits.

We also have checked the case with positive $A$ term. 
Sparticles are pulled to higher masses due to the statistical draw 
but the Higgs mass peaks at 120 GeV and less than 1\% of the scan points give correct Higgs mass. Since our main motivation for the landscape picture is the prediction of $m_h$ at its measured value, we do not 
consider experimental implications of $A_0>0$ for the remainder of the paper. 

Our goal for the present paper is to confront the panoply of recent LHC and
WIMP search experiment results with the predictions from 
the string theory landscape.

\section{Landscape predictions vs. LHC search limits}
\label{sec:lhc}

In this Section, we confront the string theory landscape 
predictions for $n=1$ with LHC sparticle search constraints and
projected reach limits.

\subsection{Landscape predictions for NUHM2 model}
\label{ssec:nuhm2}

In years past, it was common to portray collider search limits in the
$m_{1/2}$ vs. $m_0$ plane of the mSUGRA/CMSSM model \cite{kane,sugra}. 
In this model, the matter and Higgs soft mass terms are unified to a 
common GUT scale value $m_0$, where the GUT scale is defined as that scale
$m_{GUT}\simeq 2\times 10^{16}$ GeV  where the gauge couplings 
$g_1$ and $g_2$ unify. In the mSUGRA/CMSSM model, since 
$m_{H_u}^2=m_{H_d}^2\equiv m_0^2$
as an input parameter, then $\mu$ is constrained by Eq. \ref{eq:mzs}
so as to ensure the measured value of $m_Z=91.2$ GeV. 
The natural portion of parameter space using Barbieri-Giudice 
fine-tuning \cite{BG} was found to be the lowest allowed values of
$m_0$ and $m_{1/2}$ $\alt 200$ GeV \cite{AC}. 
This region is long-since excluded by LHC sparticle search constraints
which with 80 fb$^{-1}$ of integrated luminosity now require
$m_{\tg}\agt 2.25$ TeV and $m_{\tst_1}\agt 1.1$ TeV.

The value of $\mu$ in the LHC-allowed region is only natural 
$\sim 100-300$ GeV in the focus point (FP) region \cite{FP}. 
But the FP region appears only for the smaller range of $A_0$ where
$m_h$ is too low \cite{Baer:2012uya}. 
Thus, the region of mSUGRA/CMSSM parameter space
with $m_h\sim 123-127$ GeV is always highly fine-tuned \cite{Baer:2014ica}.
For this reason, we work instead first in the two-extra-parameter
non-universal Higgs model NUHM2 \cite{nuhm2} which allows
independent input values of $m_{H_u}^2$ and $m_{H_d}^2$ (since the Higgs live in
independent GUT multiplets anyway). The values of $m_{H_u}^2$ and $m_{H_d}^2$
may be traded for weak scale inputs $\mu$ and $m_A$. 
This allows us to adopt a natural value of $\mu\simeq 200$ GeV over 
all parameter space. We use Isajet 7.88 for our SUSY spectra
generation and calculation of $\Delta_{\rm EW}$ \cite{isajet}.

\begin{figure}[tbp]
\begin{center}
\includegraphics[height=0.3\textheight]{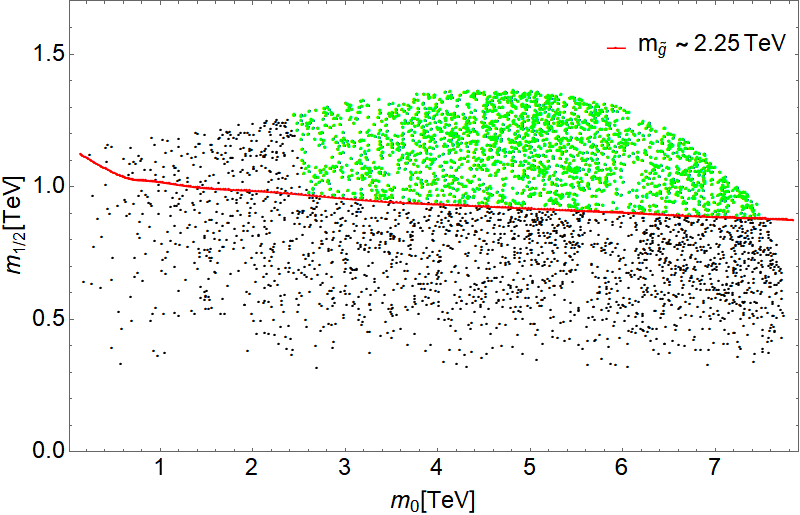}
\caption{
Locus of $n=1$ lansdscape scan points for the NUHM2 model 
with $\mu =200$ GeV in the 
$m_{1/2}$ vs. $m_0$ plane along with recent LHC Run2
constraints for 
$A_0=-1.6 m_0$, $m_A=2$ TeV and $\tan\beta =10$. The soft terms $m_0$ and 
$m_{1/2}$ are selected with $n=1$ and $\Delta_{\rm EW}<30$ so that the weak 
scale is within $\sim 4$ of its measured value. The red contour
corresponds to the LHC Run 2 limit that $m_{\tg}\agt 2.25$ TeV.
Green points are allowed by LHC Run 2 constraints (see text) while
black points are excluded by LHC Run 2. 
\label{fig:m0mhf}}
\end{center}
\end{figure}

In Fig. \ref{fig:m0mhf} we display the $m_{1/2}$ vs. $m_0$ plane of the
NUHM2 model for $\mu =200$ GeV, $\tan\beta=10$, $A_0=-1.6 m_0$ and $m_A=2$ TeV.
The soft terms $m_0$ and $m_{1/2}$ are generated randomly 
for $m_0:0\to 10$ TeV and $m_{1/2}:0.3-3$ TeV but with 
the $n=1$ increasing distribution and $\Delta_{\rm EW}<30$ so that the weak 
scale is within $\sim 4$ of its measured value. 
We see that the low $m_0$ and $m_{1/2}$ region is now sparsely populated
due to the (mild) draw of the landscape towards large soft terms.
In this plot, the density of points actually reflects the assumed 
vacuum statistics of the landscape with $n=1$. The density increases
with increasing $m_0$ and $m_{1/2}$ until the points cut off where
soft term contributions to the weak scale exceed the measured weak scale 
by a factor four ($\Delta_{\rm EW}>30$). The red line denotes the latest LHC Run 2
bound of $m_{\tg}\agt 2.25$ TeV. The green points are LHC-allowed while black 
points above the red contour at lower $m_0$ have $m_h<123$ GeV (we assume
an approximate $\pm 2$ GeV theory error in the Isajet $m_h$ calculation).
The green LHC-allowed points range up to $m_{\tg}\sim 3.5$ TeV although for
other parameter choices and moving to the NUHM3 model then gluinos can range 
as high as 6 TeV \cite{Baer:2017pba}. The most densely populated region of
parameter space remains beyond current LHC reach and it may require an 
upgrade to high energy LHC (HE-LHC with $\sqrt{s}=27$ TeV and 15 ab$^{-1}$
of integrated luminosity) to completely cover the remaining parameter space
in the gluino pair production search channel \cite{Baer:2018hpb}.

\subsection{Landscape predictions for NUHM3 model}
\label{ssec:nuhm3}

In this Subsection, we investigate landscape predictions within the more
general NUHM3 model where first/second generation matter scalars
have different soft terms from third generation matter scalars.
This sort of setup is motivated in part by investigations of the
minilandscape of heterotic string models compactified on an orbifold\cite{Nilles:2015wua}.
In these models, the first/second generation multiplets live near orbifold
fixed points and obey localized grand unification\cite{Buchmuller:2005sh}: they live in the
16-dimensional spinor reps of $SO(10)$. In contrast, the third generation
matter scalars, Higgs multiplets and gauginos live more in the bulk of the 
compactified orbifold and hence live in the usual SM split multiplets.

In this section, we will scan over soft parameters as $m_{\rm soft}^{n=1}$ with
the following ranges:
\begin{itemize}
\item $m_0(1,2): 0-55$ TeV,
\item $m_0(3): 0-20$ TeV,
\item $m_{1/2}:0-3.2$ TeV,
\item $-A_0: 0-25$ TeV and
\item $m_A:0-10$ TeV,
\end{itemize}
while we scan over $\tan\beta: 3-60$ and $\mu :100-360$ GeV uniformly
since $\tan\beta$ is not dimensional and $\mu$ arises from our assumed solution
to the SUSY $\mu$ problem. 
The lower limit of the $\mu$ term comes from the LEP2 limit 
on the lightest chargino mass, $\tw_1 >103.5$ GeV. 
We again require an appropriate EWSB and further require no contributions of 
Eq. \ref{eq:mzs} to the weak scale to exceed a factor four 
({\it i.e.} $\Delta_{\rm EW}<30$).

\begin{figure}
\centering
\begin{subfigure}[t]{0.47\textwidth}
  \centering
  \includegraphics[width=1.1\linewidth]{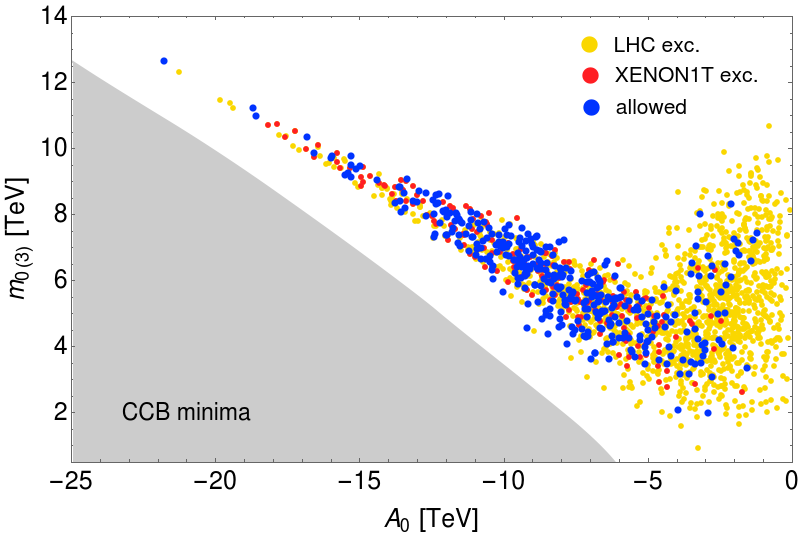}
  \caption{We show regions of $\Delta_{\rm EW}>30$ (blank) 
and CCB minima (lower left) in the scalar potential. 
The blue points are LHC Run 2 and DM-allowed.}
  \label{fig:m03A0}
\end{subfigure}%
\quad \quad
\begin{subfigure}[t]{0.47\textwidth}
  \centering
  \includegraphics[width=1.1\linewidth]{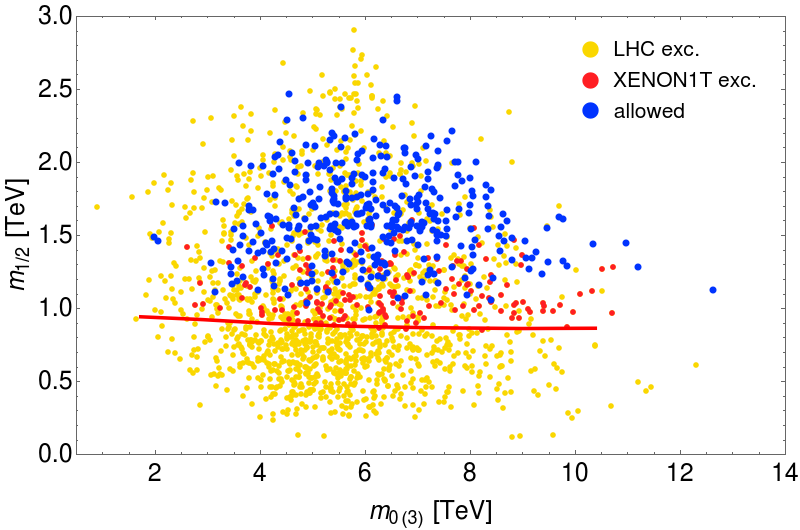}
  \caption{LHC Run 2 limit for $m_{\tg}>2.25$ TeV is shown by the red contour.}
  \label{fig:m03mhf}
\end{subfigure}
\caption{Locus of $n=1$ landscape scan points in the (a) $m_0(3)$ vs. 
$A_0$ and (b) $m_{1/2}$ vs. $m_0(3)$ planes for the NUHM3 model with $\mu =100-360$ GeV.}
\label{fig:d1}
\end{figure}

\begin{figure}
\centering
\begin{subfigure}[t]{0.47\textwidth}
  \centering
  \includegraphics[width=1.1\linewidth]{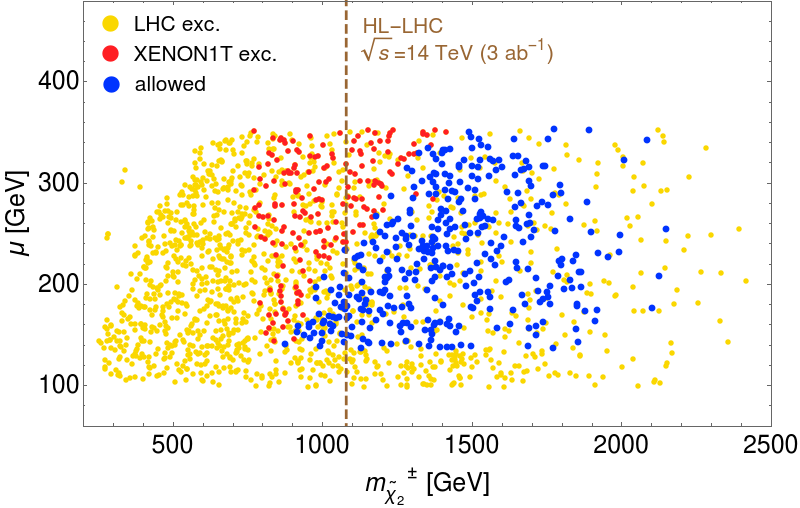}
  \caption{}
  \label{fig:mw2mu}
\end{subfigure}%
\quad \quad
\begin{subfigure}[t]{0.47\textwidth}
  \centering
  \includegraphics[width=1.1\linewidth]{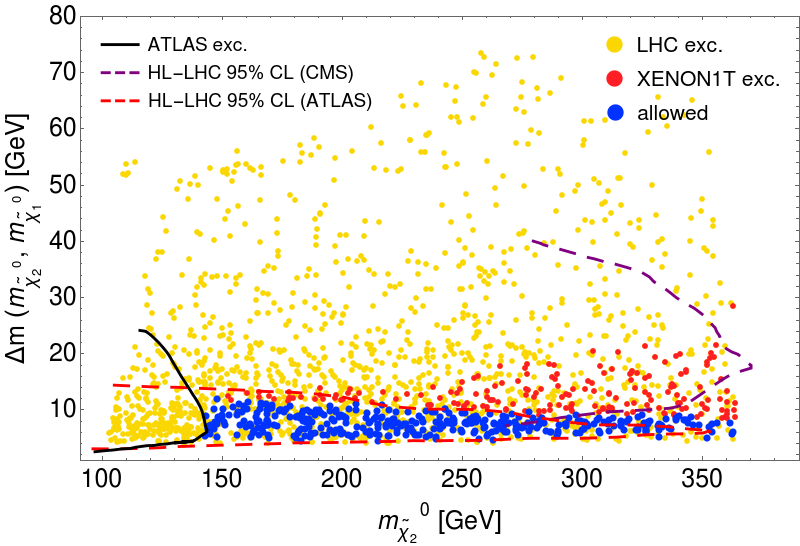}
  \caption{}
  \label{fig:dmzmz1}
\end{subfigure}
\caption{Locus of $n=1$ landscape scan points for the NUHM3 model 
with $\mu =100-360$ GeV in the (a) $\mu$ vs. 
$m_{\tw_2}$ and (b) $m_{\tz_2}-m_{\tz_1}$ vs. $m_{\tz_2}$ planes versus projected HL-LHC 95\% CL search limits.}
\label{fig:d3}
\end{figure}

Our first results are shown in Fig. \ref{fig:m03A0} where we show scan points
in the $m_0(3)$ vs. $A_0$ plane. Here we divide our scan points into three sets.
Yellow points are excluded by recent LHC Run 2 search limits:
\begin{itemize}
\item $m_{\tg}\agt 2.25$ TeV for $\tg \to t \bar{t} + \tz_1$ \cite{ATLAS:2018yhd},
\item $m_{\tst_1}\agt 1.1$ TeV for $\tilde{t} \to t^{(*)} + \tz_1$ \cite{Sirunyan:2017xse},
\item bounds from $H/A\to \tau^+\tau^-$ in the $\tan \beta$ vs. $m_A$ plane \cite{Sirunyan:2018zut},
\item higgsino pair production \cite{Aaboud:2017leg}: 
points are beyond the recent LHC soft dilepton+jets$+\eslt$ constraints
(shown later in Fig. \ref{fig:dmzmz1}),
\item $m_h=125 \pm 2$ GeV (to account for theory error of the calculation)
\end{itemize}
The blue shaded points have acceptable vacua and obey both LHC and WIMP search constraints.
The red points are LHC-allowed but excluded by recent XENON1T 
spin-independent (SI) direct WIMP detection (DD) searches \cite{Aprile:2018dbl} 
(see later Fig. \ref{fig:si}).
The main lesson from Fig. \ref{fig:m03A0} is that the acceptable points lie in
a very restricted regions where $m_0(3)$ and $-A_0$ are correlated: if
$A_0$ gets too large (negative), then the model is forced into CCB minima
so the gray region is disallowed. Likewise, if for fixed $A_0$ then
$m_0(3)$ gets too large, then third generation contributions to the weak scale
$\Sigma_u^u(\tst_{1,2})$ exceed the measured weak scale by over a 
factor four (blank region).

In Fig. \ref{fig:m03mhf} we show the $m_{1/2}$ vs. $m_0(3)$ soft term plane.
The LHC Run 2 requirement that $m_{\tg} \gtrsim 2.25$ TeV is shown by the red contour.
There are plenty of surviving landscape scan points with $m_{1/2}$ ranging 
from $1-2.5$ TeV. The upper range of allowed $m_{1/2}$ values correspond
to values of $m_{\tg}$ over 6 TeV. Few scan points exist for
$m_0(3)\alt 2$ TeV since the $n=1$ scan prefers linearly increasing soft terms. Few points also exist for $m_0(3)\agt 12$ TeV
since these points would give too large $\Sigma_u^u(\tst_{1,2})$ contributions
to the weak scale.

\begin{figure}
\centering
\begin{subfigure}[t]{0.47\textwidth}
  \centering
  \includegraphics[width=1.1\linewidth]{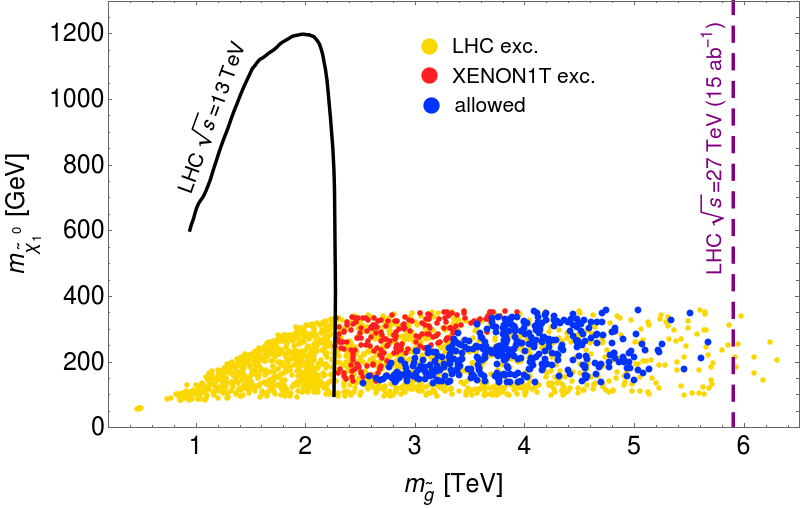}
  \caption{}
  \label{fig:mglmz1}
\end{subfigure}%
\quad \quad
\begin{subfigure}[t]{0.47\textwidth}
  \centering
  \includegraphics[width=1.1\linewidth]{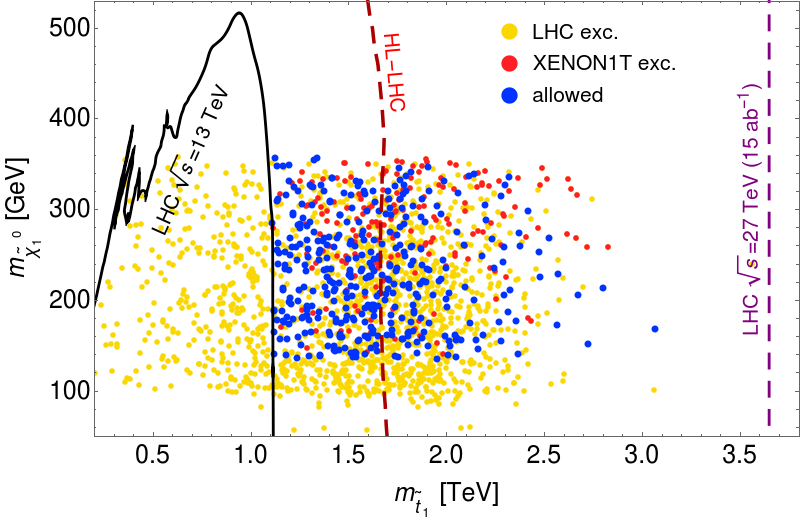}
  \caption{}
  \label{fig:mt1mz1}
\end{subfigure}
\caption{Locus of $n=1$ landscape scan points for the 
NUHM3 model with $\mu =100-360$ GeV in the (a) $m_{\tz_1}$ vs. $m_{\tg}$ 
and (b) $m_{\tz_1}$ vs. $m_{\tst_1}$ planes versus recent LHC Run2 constraints.}
\label{fig:d2}
\end{figure}

In Fig. \ref{fig:mglmz1}, we show our $n=1$ landscape scan points in the
$m_{\tz_1}$ vs. $m_{\tg}$ plane which is the usual simplified model plane
in which LHC gluino searches are usually presented.
The current LHC Run 2 exclusion contour based on 80 fb$^{-1}$ of 
integrated luminosity is shown as the black contour. 
It is also interesting that the XENON1T dark matter search excludes 
significant regions of the lighter LSP masses for gluino masses 
of order $2-3.5$ TeV. HL-LHC will be able to cover points with gluinos 
only up to 2.8 TeV via the gluino pair production channel \cite{mgluino}. 
We also show the recently computed 95\% CL HE-LHC projected 
search limit for $\sqrt{s}=27$ TeV and  15 ab$^{-1}$ which reaches to
$m_{\tg}\sim 6$ TeV. 
Evidently a complete examination of $n=1$ landscape points 
in the $\tg\tg$ search channels will require HE-LHC.

In Fig. \ref{fig:mt1mz1}, we show $n=1$ landscape points in the 
$m_{\tz_1}$ vs. $m_{\tst_1}$ simplified model plane. The current LHC Run 2 search
limits are shown as the black contour. There is a high density of LHC-allowed 
(blue) points extending from $m_{\tst_1}\sim 1.1$ to $2.7$ TeV. 
The projected reach of HL-LHC with $\sqrt{s}=14$ TeV and 3 ab$^{-1}$ extends to $m_{\tst_1}\sim 1.7$ TeV
and covers perhaps the greatest density of blue points. Nonetheless, it will
require an upgrade to HE-LHC to cover the complete set of $n=1$ 
landscape points \cite{Baer:2018hpb}.
\begin{figure}
\centering
\begin{minipage}[t]{.47\textwidth}
  \centering
  \includegraphics[width=1\linewidth]{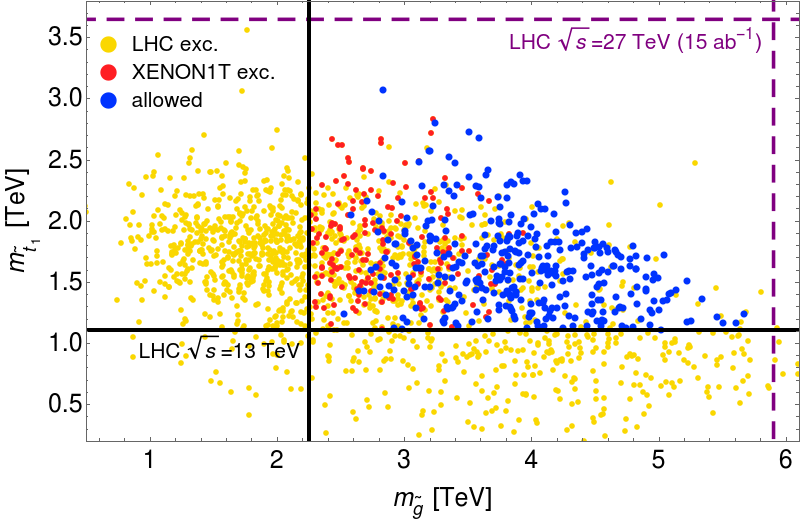}
  \captionof{figure}{Locus of $n=1$ landscape scan points for the 
NUHM3 model with $\mu =100-360$ GeV in the 
  $m_{\tst_1}$ vs. $m_{\tg}$ plane versus recent LHC Run2 constraints (black) and 
  projected HE-LHC 95\% CL reach contours (purple-dashed).}
  \label{fig:mt1mgl}
\end{minipage}%
\quad \quad
\begin{minipage}[t]{.47\textwidth}
  \centering
  \includegraphics[width=1\linewidth]{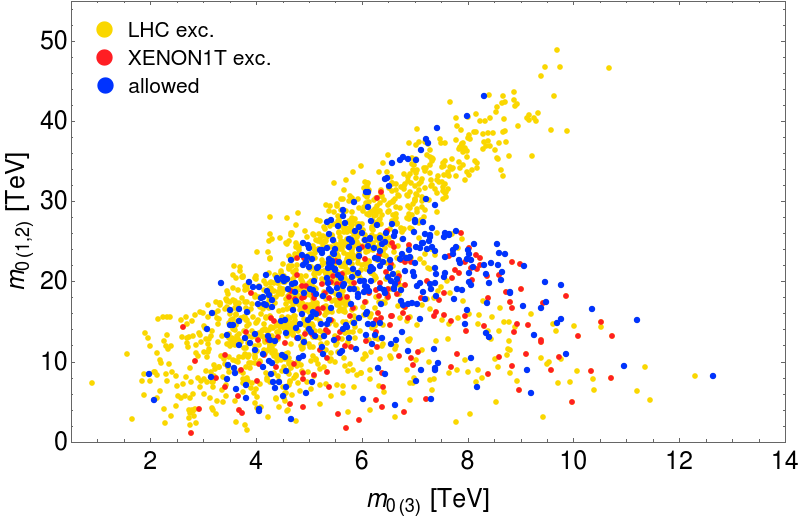}
  \captionof{figure}{Locus of $n=1$ landscape scan points for the NUHM3 model
 in the  $m_0(1,2)$ vs. $m_0(3)$ plane for $\mu =100-360$ GeV.}
  \label{fig:m012_m03}
\end{minipage}
\end{figure}

In Fig. \ref{fig:mt1mgl}, we show the $n=1$ landscape points in the 
$m_{\tg}$ vs. $m_{\tst_1}$ plane. Of note here is that the points 
with the largest values of $m_{\tg}$ have the smaller range of $m_{\tst_1}$ and
vice-versa. Thus, if somehow for instance gluinos were able to escape LHC 
detection on account of them being too heavy, the top squarks would surely be
seen (and vice-versa). A complete coverage of the $n=1$ landscape 
parameter space will require HE-LHC with 15 ab$^{-1}$ of integrated 
luminosity.

In Fig. \ref{fig:m012_m03}, we show the $m_0(1,2)$ vs. $m_0(3)$ plane
of the NUHM3 model for the $n=1$ landscape. The important lesson from this plot is that first/second generation matter scalar soft terms tend to inhabit the
10-30 TeV range whilst third generation matter scalar soft terms lie 
typically below 10 TeV. RG and mixing effects then cause the third generation 
squarks/sleptons to lie in the few TeV range (so their loop-suppressed
contributions $\Sigma_u^u$ to the weak scale are small) while first/second
generation squarks and slepton (with mass $m_{\tq,\tell}\sim m_0(1,2)$)
lie well beyond even HE-LHC reach and offer at least a partial 
decoupling solution to the SUSY flavor and CP problems. 
The reason they can be so heavy is that the first/second generation sfermion
contributions to the weak scale are $D$-term contributions which largely
cancel \cite{Baer:2013jla}.

Another important LHC search plane is the $\mu$ vs. $m_{\tw_2}$ plane
This plane is important for presenting search limits from same-sign diboson
(SSdB) production arising from wino pair production in SUSY models with
light higgsinos. The reaction is $pp\to\tw_2 \tz_4$ where
$\tw_2 \to W^\pm\tz_{1,2}$ while $\tz_4\to W^\mp\tw_1 $ so that
half the time one arrives at a final state with two same-sign $W$ bosons 
plus large $\eslt$. 
For leptonically-decaying $W$ bosons, then the final state consists of 
a same-sign dilepton $+\eslt$ signature which is relatively jet free- in 
contrast to same-sign dileptons originating from gluino and squark
pair production.
The same-sign channel has rather tiny SM backgrounds \cite{ssdb1,rns@lhc,ssdb2}.
So far, no search results have been presented by ATLAS or CMS.
From Fig. \ref{fig:mw2mu}, we see that LHC-allowed points only begin
at wino masses $m_{\tw_2}\sim 800$ GeV and then extend out to
$m_{\tw_2}\sim 2300$ GeV. This is to be compared with the 
projected HL-LHC 95\% CL search limit which is the brown contour reaching to 
$m_{\tw_2}\sim 1100$ GeV \cite{ssdb2}. 
Projected search limits for HE-LHC in the SSdB  channel have yet to be computed.

In Fig. \ref{fig:dmzmz1}, we show the $n=1$ landscape points in the
$\Delta m\equiv m_{\tz_2}-m_{\tz_1}$ vs. $m_{\tz_2}$ plane. 
This plane is important for the light higgsino pair production
searches $pp\to \tz_2\tz_1$ (and $\tz_2\tw_1$) where $\tz_2\to\ell^+\ell^-\tz_1$
giving rise to a soft opposite-sign dilepton pair whose invariant mass is 
bounded by $m_{\tz_2}-m_{\tz_1}$ \cite{SDLJMET}. To trigger on such events, it seems necessary 
to require a hard jet radiation from the initial state against which the 
soft dileptons can recoil. Recent search limits have been presented by
both ATLAS \cite{Aaboud:2017leg} and CMS \cite{Sirunyan:2018iwl}. 
We show as a black contour the recent ATLAS limit. 
An important point of the $n=1$ landscape is that it favors 
heavier gauginos while $\mu$ must not be too far from the weak scale.
The combination squeezes down the inter-higgsino mass gap $m_{\tz_2}-m_{\tz_1}$
so that in this case all the LHC-allowed points have $\Delta m\alt 10$ GeV.
We also show recently computed projected HL-LHC 95\% CL reach contours \cite{CidVidal:2018eel}.
The ATLAS contour has focussed on the small mass gap region and appears
to cover nearly all parameter space. This has important implications
for how SUSY is likely to be revealed at LHC. Gluinos and top squarks
are expected to be drawn to large values, possibly beyond HL-LHC reach.
The soft dilepton plus jet$+\eslt$ channel (SDLJMET) is the 
only channel that seems to be (nearly) completely covered by HL-LHC. 
Thus, we would expect a SUSY signal in this channel to emerge slowly 
but conclusively during the next 15 years as LHC acquires its 
full complement of 3 ab$^{-1}$ of integrated luminosity!
\begin{figure}[tbp]
\begin{center}
\includegraphics[height=0.3\textheight]{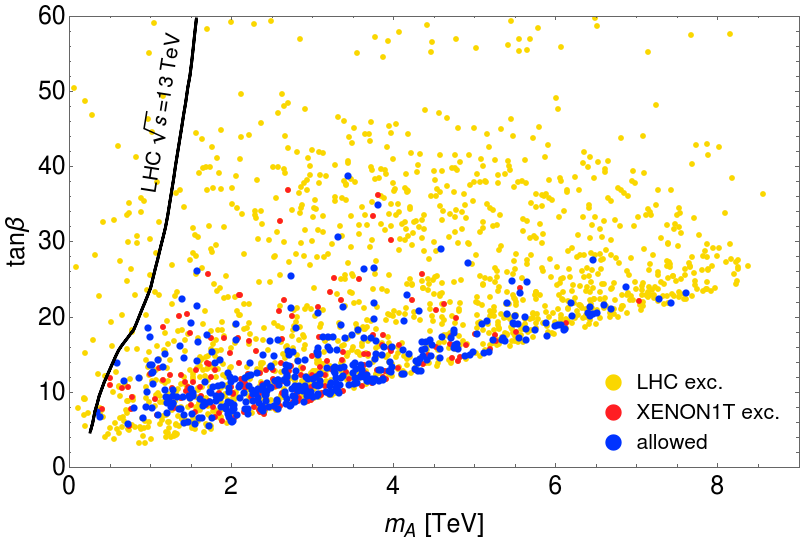}
\caption{
Locus of $n=1$ landscape scan points for the NUHM3 model with 
$\mu =100-360$ GeV in the $\tan\beta$ vs. 
$m_A$ plane versus recent LHC Run2 constraints. Blue points are LHC Run 2 
and DM-allowed while red points are LHC-safe but 
excluded by XENON1T WIMP search limits.
\label{fig:tanBmA}}
\end{center}
\end{figure}

A final natural SUSY search channel occurs in the Higgs sector by
looking for $pp\to A,H\to\tau^+\tau^-$ events. These searches are typically 
presented in the $\tan\beta$ vs. $m_A$ plane which we show in 
Fig. \ref{fig:tanBmA}. We also show the recent LHC excluded region
as the black contour\cite{Sirunyan:2018zut}. 
This latter contour assumes only SM decay modes
for the $A$ and $H$ but for the landscape then the decay modes
$H,A\to higgsinos$ should almost always be open as well
(and might lead to $4\ell+\eslt$ signatures \cite{Baer:1994fx}).
The added $H,A\to higgsinos$ decay modes hardly affect the search limits
since a diminution of Higgs to SM branching fractions can be offset by
increasing the Higgs production cross sections by moving to 
somewhat larger $\tan\beta$ \cite{Bae:2015nva}.
From the plot, we see that the LHC-allowed points are typically 
well beyond the current LHC reach limits and the greatest density 
populates the region of $m_A\sim 2-5$ TeV and at lower $\tan\beta$ values
where $b\bar{b}$ fusion contributions to the production cross section 
are not so big. 
Thus, we see only a small  likelihood of a signal emerging in this channel 
at LHC.

\section{Landscape predictions vs. WIMP DM search limits}
\label{sec:wimp}

In $n=1$ landscape SUSY, we expect soft terms to be drawn to large values 
whilst the $\mu$ term is not too far from the weak scale. This results in a 
Little Hierarchy (LH) with $\mu\ll m_{soft}$ which turns out to be 
non-problematic. 
In such a scenario, then the lightest SUSY particle is the lightest 
higgsino-like neutralino  with mass $m_{\tz_1}\sim 90-360$ GeV. These
natural higgsino-like WIMPs are thermally underproduced as dark matter, 
which may be a reason why they had not been considered much previous to 
2011 \cite{bbh,Allahverdi:2012wb,Baer:2013vpa}. 
However, one must recall that the QCD sector of 
the SM also suffers a naturalness issue in the form of the strong CP problem. 
Including an axion sector into the MSSM is thus well-motivated both for
solving the strong CP problem but also to solve the SUSY $\mu$ problem
(for a recent review, see {\it e.g.} Ref. \cite{mupaper}) via a DFSZ-type
SUSY axion. 
Thus, in natural SUSY it is expected that the DM is a WIMP-axion
admixture ({\it i.e.} two dark matter particles). 
If the natural SUSY WIMPs (with $m_{\tz_1}\alt 350$ GeV) 
were {\it all} of DM, then they would actually be excluded 
by current WIMP search constraints \cite{Baer:2018rhs}.
But if the DM is mainly axions, then there are far fewer relic WIMPs
present in the cosmos and they can escape present limits from WIMP 
search experiments. A full evaluation of mixed WIMP-axion dark matter
requires an eight-coupled Boltzmann equation evaluation which accounts
also for axino, saxion and gravitino production and decay in the 
early universe \cite{Bae:2014rfa}.
\begin{figure}[tbp]
\begin{center}
\includegraphics[height=0.3\textheight]{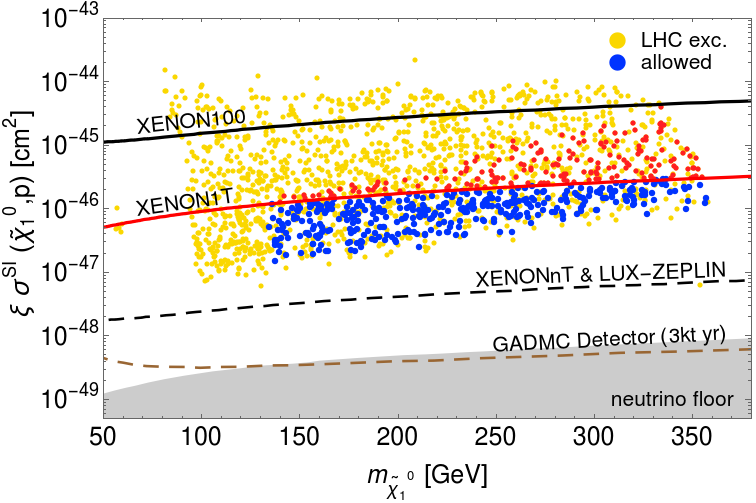}
\caption{Locus of $n=1$ landscape scan points for the NUHM3 model
in the $\xi\sigma^{SI}(\tz_1, p)$ vs. $m_{\tz_1}$ plane versus recent 
WIMP search constraints for $\mu =100-360$ GeV.
Red points are excluded by XENON1T search 
limits but not by LHC Run 2 constraints. 
Projected reaches from several future SI DD experimentes are also 
shown.
\label{fig:si}}
\end{center}
\end{figure}

We first examine WIMP search limits via ton-scale noble liquid experiments
using targets such as Xenon or Argon. 
To compare WIMP search limits to landscape projections, one must calculate
$\xi \sigma^{SI}(\tz_1, p)$ where $\xi\equiv \Omega_{\tz_1}h^2/0.12$, 
{\it i.e.} it is the fractional abundance of WIMPs in making up the totality
of dark matter. Usually this is just the WIMP thermal abundance divided by
the measured abundance although it is possible the WIMP abundance is
supplemented by non-thermal processes such as axino or saxion decay in the 
early universe.

In Fig. \ref{fig:si}, we plot the locus of $n=1$ landscape points in the
$\xi\sigma^{SI}(\tz_1, p)$ vs. $m_{\tz_1}$ plane. 
We also show current search limits from the XENON-100 experiment \cite{Aprile:2016swn}
(black contour) and the XENON1T experiment \cite{Aprile:2018dbl} (red contour). 
A subset of LHC-allowed points are already excluded, and denoted as red points.
However, the bulk of $n=1$ landscape points are still allowed, 
and extend down to an order of magnitude below present limits.
These points do not extend all the way to the neutrino floor since in 
SUSY the WIMPs couple to nucleons mainly via light Higgs exchange and this
coupling involves a production of gaugino times higgsino components\cite{Baer:2013vpa}. 
In natural SUSY, the WIMP is mainly higgsino, but with non-negligible
gaugino component (lest heavy gauginos give too large a contribution to the 
weak scale). Thus, it appears that projected seach limits from XENONnT
(multi-ton Xenon detector), LUX-ZEPLIN (LZ) \cite{LZ} and other 
multi-ton-scale detectors \cite{futureSI} should cover the entire $n=1$ landscape
parameter space, even if WIMPs comprise only a portion of the dark 
matter. 
%
%
\begin{figure}
\centering
\begin{subfigure}[t]{0.47\textwidth}
  \centering
  \includegraphics[width=1.1\linewidth]{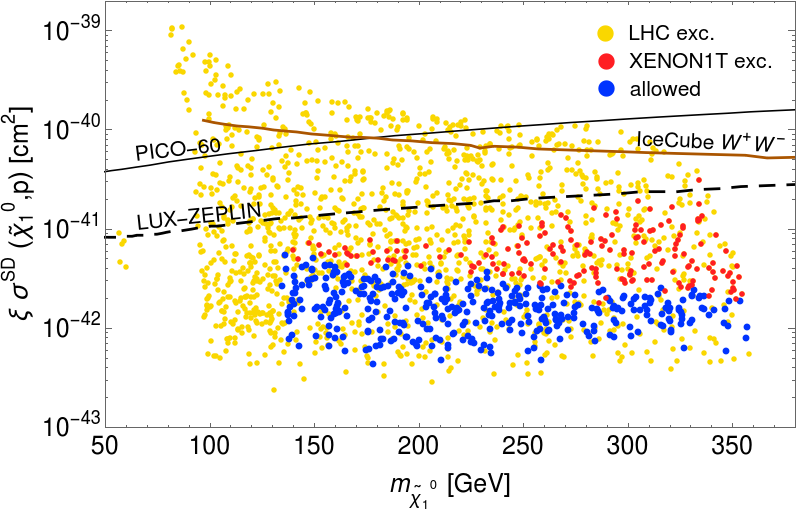}
  \caption{spin-dependent DM detection rates}
  \label{fig:sd}
\end{subfigure}%
\quad \quad
\begin{subfigure}[t]{0.47\textwidth}
  \centering
  \includegraphics[width=1.1\linewidth]{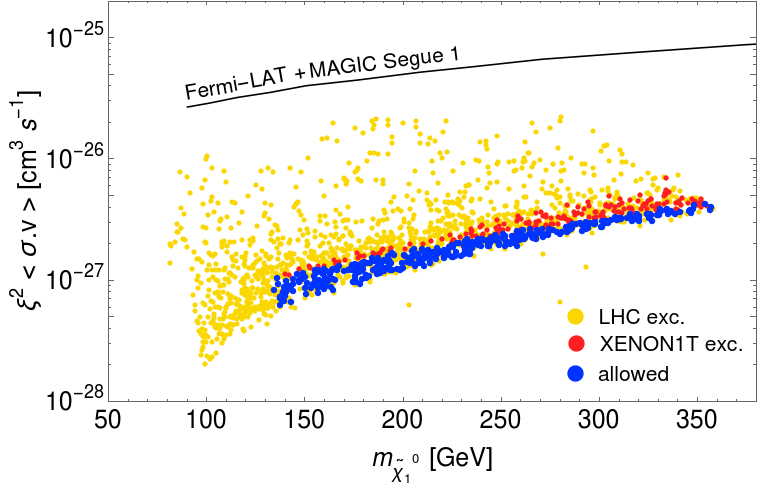}
  \caption{indirect DM detection rates}
  \label{fig:idd}
\end{subfigure}
\caption{Locus of $n=1$ landscape scan points for the 
NUHM3 model with $\mu =100-360$ GeV in the 
(a) $\xi\sigma^{SD}(\tz_1, p)$ vs. $m_{\tz_1}$ 
and (b) $\xi^2\langle\sigma v\rangle$ vs. $m_{\tz_1}$ planes versus recent WIMP search constraints.}
\label{fig:d4}
\end{figure}

In Fig. \ref{fig:sd}, we show the spin-dependent (SD) direct detection 
scattering rate $\xi\sigma^{SD}(\tz_1, p)$ vs. $m_{\tz_1}$ plane along with
projected $n=1$ landscape rates. We also show recent limits 
from the PICO-60 experiment \cite{Amole:2017dex} and IceCube in the
$W^+W^-$ annihilation mode \cite{Aartsen:2016zhm}. In this case, the 
LHC-allowed landscape points tend to lie about an order of magnitude below
the current limits. We also show the projected future reach of LZ \cite{LZ}.
Even this contour does not quite reach the expected theory region.

%

In Fig. \ref{fig:idd}, we show the indirect WIMP detection rates (IDD)
in the $\xi^2\langle\sigma\cdot v\rangle$ vs. $m_{\tz_1}$ plane, where
$\xi^2$ is required since these signals arise from WIMP-WIMP
annihilation in the cosmos and is thus suppressed by the fractional
WIMP abundance squared. We show also the recent Fermi-LAT+Magic
limits from observation of dwarf spheroidal galaxy Segue-1 \cite{Ahnen:2016qkx}. 
The current limit is over an order of magnitude above the expected
LHC-allowed points from the $n=1$ landscape. A signal from the 
IDD search channel would point to non-thermal production of WIMPs 
($i.e.$ from decays of axinos/saxions) 
in addition to the thermally-produced neutralinos.
%

\section{Conclusions}
\label{sec:conclude}

Rather general arguments regarding the statistics of the landscape of 
flux vacua in string theory point to a statistical draw towards large soft
SUSY breaking terms governed by a power law $m_{soft}^n$ where
$n=2n_F+n_D-1$ involves the number of $F$ and $D$ term SUSY breaking
fields in a (possibly) complicated hidden sector. With only the draw towards large soft terms, one would expect a huge value for the weak scale since
$m_{weak}$ is determined by the visible sector soft terms and the SUSY $\mu$ 
parameter. A huge weak scale would mean highly suppressed weak 
interactions and huge particle masses which would likely lead to a 
non life-supporting universe. Agrawal {\it et al.} calculated that an
increase in $m_{weak}$ by a factor of $\sim 3$ would lead to a non-livable
universe. Therefore, we have tempered the statistical draw of the landscape
to large soft terms with the anthropic requirement of a weak scale 
no more than (conservatively) four times its measured value. 
For a fixed natural value of $\mu$
(arising from some solution to the SUSY $\mu$ problem such as hybrid CCK
which also solves the strong CP problem and introduces axionic along with 
higgsino-like WIMP dark matter\cite{Baer:2018avn}), 
then one may implement random scans over power-law
increasing soft terms tempered by an anthropic requirement of the 
weak scale not too far from $m_{weak}\sim 100$ GeV.
The cases for $n=1$ and 2 lead to a landscape probability distribution 
for the light Higgs mass which peaks around $\sim 125$ GeV. Also, 
most superparticle masses are pulled to large values beyond LHC reach.
In this case, {\it LHC sparticle and WIMP dark matter search experiments are
seeing exactly that which is expected from the $n=1,2$ landscape}: 
a Higgs mass of 125 GeV and no sign yet of sparticles.

Our goal in this paper was a practical one: place the $n=1$ landscape
statistical predictions on the same plots that LHC sparticle 
and WIMP dark matter
search experiments use in order to assess where sparticle and WIMP
masses might be located relative to present and future search limits.
In these sorts of plots, the density of points actually has meaning 
since it would reflect the assumed statistics of string theory vacua 
in a fertile patch which includes the MSSM as the weak scale effective theory. 

Our findings are that strongly interacting sparticles 
$\tg$ and $\tst_1$ are likely to
lie beyond present LHC search bounds and possibly beyond HL-LHC 
projected search limits. 
It may require an upgrade to HE-LHC to cover the entire $n=1$ landscape 
parameter space in the $\tg\tg$ and $\tst_1\tst_1^*$ modes 
as manifested in the context of the natural NUHM3 SUSY model. 
Also, the SSdB signal may or may not be detected by HL-LHC
and the $A,H\to\tau^+\tau^-$ signals are likely to lie well beyond the reach 
of LHC upgrades. However, the SDLJMET signal arising from higgsino pair
production offers a search channel wherein HL-LHC may cover the 
entire projected parameter space. 
In this case, we would expect a 
SUSY signal to emerge slowly but conclusively by ATLAS and CMS as more and more
integrated luminosity accrues.

Regarding WIMP searches, it appears that a full complement of data from 
multi-ton noble liquid SI direct detection experiments should cover 
the entire $n=1$, NUHM3 landscape parameter space. 
This can occur even though in natural SUSY the higgsino-like WIMPs make up 
only a portion of dark matter with SUSY DFSZ axions making up the remainder.
Meanwhile, even upgraded SD detectors may well fall short of covering the
portion of parameter space occupied by the $n=1$ landscape model.
It appears that IDD WIMP search experiments will also have a hard time accessing the full parameter space since now the expected signal rates are diminished 
by the square of the fractional WIMP abundance.
The search for SUSY DFSZ axions, addressed in Ref. \cite{Bae:2017hlp}, is also difficult 
due to the presence of higgsinos circulating in the $a\gamma\gamma$ 
coupling diagram: they tend to suppress the DFSZ axion coupling to much lower
values than are typically displayed in axion search result plots.

To summarize:
\begin{itemize}
\item The $n=1$ landscape statistics predict 
$m_h\sim 125$ GeV with superparticles beyond the reach of present LHC and 
WIMP DD experiments.
\item LHC signals in SDLJMET channel should emerge slowly as LHC attains
higher and higher integrated luminosity. A signal should likely be seen 
with 3 ab$^{-1}$ at HL-LHC if not sooner. 
Signals in other channels such as
$\tg\tg$, $\tst_1\tst_1^*$ and SSdB may emerge at HL-LHC if we are lucky
but otherwise may require an upgrade to HE-LHC.
\item WIMP detection signals should emerge in SI DD experiments using multi-ton
noble liquids. Signals in SD DD or IDD are much less likely to emerge in the near future.
\end{itemize}

{\it Acknowledgements:} We thank Art McDonald for discussions on future WIMP detectors.
This work was supported in part by the US Department of Energy, Office
of High Energy Physics. The computing for this project was performed at the OU Supercomputing Center
for Education \& Research (OSCER) at the University of Oklahoma (OU).


%
\end{document}